\documentclass[manuscript]{aastex}

\usepackage{natbib}
\newcommand{\sxtot}{\sigma_{tot}}
\newcommand{\sxgas}{\sigma_{gas}}
\newcommand{\sxdust}{\sigma_{dust}}

\newcommand{\fb}{f_b(E)}

\newcommand{\nh}{n_H}

\shorttitle{Photoelectric cross-sections in protoplanetary disks}
\shortauthors{Bethell \& Bergin}

\bibstyle{apj.bst}

\begin{document}


\title{Photoelectric cross-sections of gas and dust in protoplanetary disks}


\author{T. J. Bethell and Edwin A. Bergin}
\affil{Astronomy Department, University of Michigan, Ann Arbor, MI 48109}

\email{tbethell@umich.edu}


%\altaffiltext{1}{Visiting Astronomer, Cerro Tololo Inter-American Observatory.
%CTIO is operated by AURA, Inc.\ under contract to the National Science


\begin{abstract}
We provide simple polynomial fits to the X-ray photoelectric cross-sections 
($0.03 < E < 10$keV) for mixtures of gas and dust found in protoplanetary disks.  Using the solar elemental abundances of Asplund et al. (2009) we treat the gas and dust components separately, facilitating the further exploration evolutionary processes such as grain settling and gain growth. We find that blanketing due to advanced grain-growth $(a_{max} > 1\mu m)$ can reduce the X-ray opacity of dust appreciably at $E_X \sim 1$keV, coincident with the peak of typical T Tauri X-ray spectra.  However, the reduction of dust opacity by dust settling, which is known to occur in protoplanetary disks, is probably a more significant effect. The absorption of 1-10keV X-rays is dominated by gas opacity once the dust abundance has been reduced to about 1\% of its diffuse interstellar value. The gas disk establishes a floor to the opacity at which point X-ray transport becomes insensitive to further dust evolution. Our choice of fitting function follows that of Morrison \& McCammon (1983), providing a degree of backward-compatibility. 

\end{abstract}


%\keywords{}

%% and \citet commands to identify citations.  

\section{Introduction\label{sec:intro}}

%Since the early days of X-ray astronomy there has been a need to understand how 
%intervening interstellar material alters the X-ray spectra of Galactic and extra-galactic 
%objects \citep{Brandt:2005mb,Ueda:2003if}. Furthermore, the X-ray absorption 
%and scattering properties of material plays an important role when studying the internal 
%physics of energetic astrophysical systems \citep{Beiersdorfer:2003qq,Feigelson:1999fc,Tozzi:2006zh}. 
The evaluation of interstellar X-ray cross-sections stretches back over four decades \citep[e.g.][]{Brown:1970ys} and is essential for interpreting X-ray observations subject to absorption by intervening material.  Today, the evaluation and manipulation of highly detailed X-ray photoelectric cross-sections for arbitrary elemental abundances is made possible by powerful publicly available codes\footnote{These codes include XSPEC \citep[][http://heasarc.nasa.gov/xanadu/xspec/]{Arnaud:1996zt}, SHERPA \citep[][http://cxc.harvard.edu/sherpa/]{Freeman:2001ys}, and ISIS \citep[][http://space.mit.edu/CXC/isis/]{Houck:2000vn}}. Of some historical importance is the paper by \citet[][hereafter MM83]{Morrison:1983fx} in which the authors present the photoelectric cross-section using a simple piece-wise quadratic polynomial fitted over discrete energy intervals (spanning a total range $0.03 < E < 10$keV). The main benefit of using a fitting formula is that it is an expeditious way to incorporate the cross-sections into a computer code.  The basic limitation of this approach is that elemental abundances become Ôlocked in.Õ This was relaxed somewhat in the follow-up paper by \citet{Balucinska-Church:1992kn}, while \citet[][hereafter W00]{Wilms:2000zv} made adjustments for the H$_2$ molecule and included more accurate estimates of elemental abundances. Both these works treated the individual elemental abundances as adjustable parameters, increasing the versatility of the results whilst increasing the complexity of the implementation.  

Central to the piece-wise construction of MM83 is the fact that - despite their low cosmic abundances relative to H and He - the metals (C, O, Ne, Mg, etc.) contribute significantly to the X-ray opacity at energies above their respective K-shell photoelectric thresholds ($0.28$keV for C,  $0.53$keV for O,  $0.87$keV for Ne, $1.3$keV for Mg, for example). These are the elements which are most readily incorporated into dust grains, although under typical interstellar conditions significant fractions of C and O (mostly in the form of CO), and Noble elements, persist in the gas phase. As such, the deposition of X-ray energy ($E>0.3$keV) occurs through absorption by both gas and dust.  The relative importance of gas and dust is highly energy dependent.  At low energies, $E<1$keV, only a handful of metals have sufficiently low K-shell thresholds that they can add significantly to the opacity floor set by H+He.  In this case most $E\sim1$keV X-rays will be processed by the gas.  While a greater variety of elements are susceptible to K-shell ionization at higher energies ($E\sim 10$keV), the strong $\sim E^{-3}$ dependence of the photoelectric cross-section means that the total cross-section drops rapidly with increasing energy.  Here heavy metals such as Fe dominate the photoelectric cross-section, implying that 10keV X-rays will likely be processed by grains.
 
It is well established observationally that young pre-main sequence (T Tauri) stars are X-ray luminous with steady X-ray spectra typically peaking at energies of $E\sim1-2$keV \citep{Feigelson:2005dw}.  In addition, there is considerable variability due to the eruption of stellar flares which generate a relatively hard X-ray component, briefly extending the X-ray spectrum to beyond $10$keV.  In order to understand how X-rays interact with the disk it is necessary to consider how the evolution of protoplanetary disk material might affect the total gas+dust photoelectric cross-section. In this paper we extend the approach of MM83 by considering the effect on the X-ray photoelectric cross-section induced by two fundamental grain processes believed to occur in protoplanetary disks; dust settling and grain-growth. 

Interstellar dust grains become optically thick to X-rays ($E\sim1$keV) once they grow to radii of $a\ge0.5\mu$m. At this point the interior atoms will be shielded from the impinging X-rays by the layers of atoms closer to the grain surface: the interior atoms do not see the full radiation Þeld and so do not contribute equally to the X-ray cross-section. This effect is referred to as self-blanketing \citep[][W00]{Fireman:1974ay}, rendering dust less efficient at absorbing X-rays. This reduction in absorbing efficiency is encapsulated in the self-blanketing parameter $\fb$. As grains continue to grow to larger sizes (the precise size is highly energy dependent) essentially all X-rays incident upon them are absorbed (the regime of geometric optics). However, even the largest dust grains in the diffuse ISM \citep[$a_{max}\sim0.25$mm,][hereafter MRN]{Mathis:1977bd} are only marginally thick to 1keV X-rays, whereas the inferred grain-size distribution in protoplanetary disks can extend to signiÞcantly larger sizes \citep[$a_{max} > 1$mm,][]{Throop:2001nx,Wilner:2005hs,Lommen:2009uf}, making them strongly self-blanketing in the $E=1-10$keV range. Whether large grains actually contribute to the absorption of stellar X-rays depends largely upon whether they form part of the gas-dust mixture in the X-ray irradiated surface layers. 

In a steady protoplanetary disk the vertical component of the stellar gravitational Þeld, combined with the differential drift between the ballistic grains and partially centripetally supported gas, generates friction between the gas and dust \citep{Weidenschilling:1993kc}. The subsequent drag forces lead to vertical settling of dust toward the disk midplane, occurring on a timescale of order $10^5$years. This is short compared to typical disk lifetimes ($> 10^6$ yrs), suggesting that disks exhibit a dust-poor atmosphere overlying a thin, dust-rich midplane \citep{Dullemond:2004kh}. Following the notation of  \citet{DAlessio:2006cq} the extent of grain removal from the upper disk layers is represented by the parameter $\epsilon$, which we simply define as the local dust-gas mass ratio relative to that in the ISM (i.e. in the ISM $\epsilon=1$). While dust settling occurs aggressively in typical T Tauri systems \citep[the median value in Taurus is $\epsilon\sim0.01$,][]{Furlan:2006ss}, the modeling of infrared spectral energy distributions suggests that there is a continual replenishment and recirculation of small grains in the upper layers of the disk \citep{Dullemond:2008pt}. 

In this paper we return to the approach adopted by MM83, providing polynomial fits to the X-ray cross-sections using the most recent estimates for the solar elemental abundances provided by \citet{Asplund:2009jl}. By explicitly splitting the total X-ray cross-section into gas and dust components it is possible to explore the effects of dust-speciÞc processes (grain growth and settling). While the potential effect of grain-growth was mentioned in the aforementioned papers, it was shown to be of little quantitative importance for grains in the diffuse interstellar medium (typically affecting the cross-sections at the few percent level). In contrast, these effects are expected to become important in protostellar and protoplanetary systems where the nature of the gas-dust mixture departs signiÞcantly from that found in the interstellar medium \citep{Beckwith:2000gb}.  Such results may facilitate the modeling of X-ray transport and energy deposition \citep{Maloney:1996kl,Igea:1999jo,Nomura:2007cr}; resolving the various roles of X-rays in relation to other sources of energy \citep[e.g. cosmic-rays and ultraviolet irradiation,][]{Glassgold:2004yo,Ercolano:2008rq,Ercolano:2009hq,Owen:2010kx}; and interpreting the observed absorption of X-rays that sample disk material \citep{Gudel:2008zr,Kastner:2005ly}.
We present separate polynomial fits for three components; 
\begin{enumerate}
\item A \textit{gas} composed of hydrogen and helium. 
\item A \textit{gas} composed of hydrogen, helium, the noble gases, and some fraction of the carbon 
and oxygen (required to make gas-phase CO). 
\item \textit{Dust grains} containing all the elements heavier than helium, except the noble gases and a fraction of the carbon and oxygen.
\end{enumerate}
We regard Component 1 as a basal reference appropriate for chemically depleted gas. Component 2 is more typical of the \textit{gas} in classical T Tauri disks. Component 3 is the  \textit{dust}, and it is to this that we apply our grain-specific effects.   Components 2 and 3 together constitute disk material.  The remainder of this paper is structured as follows. In Section 2 we describe the fitting formula and the modifications required to include the grain-speciÞc physics. The composition of the separate gas and dust components are given in terms of the latest solar abundances. In Section 3 the resulting X-ray cross-section fits for the various components are provided both in graphical and tabulated forms. The paper concludes with Section 4, providing a summary of the results.

\section{X-ray cross-section fits\label{sec:fits}}

In this section we explicitly separate the gas and dust components and provide 
polynomial fits to their cross-sections. To enable backward compatibility with MM83 
we have adopted the same fitting function and energy ranges. The X-ray photoelectric 
cross-section per H nucleus, $\sigma(E)$, is described by the piecewise polynomial fitting function, 

\begin{equation}
\sigma(E)=10^{-24}\times\left(c_{0}+c_{1}E+c_{2}E^{2}\right)E^{-3}\textrm{cm\ensuremath{^{2}}}.\label{eq:fn_fitting}\end{equation}where $E$ is the X-ray energy and  $c_{0}$, $c_{1}$, and $c_{2}$ are the coefficients to be found. We extend Equation \ref{eq:fn_fitting} by decomposing the medium into a mixture of gas and dust \citep[e.g.][]{Fireman:1974ay}. Initially we might split the total cross-section into two contributions so that $\sxtot=\sxgas+\sxdust$ (units cm$^2$ per H nucleus). From the point of view of computing X-ray opacities, the relatively small dust grains in the diffuse ISM can be treated as though 
their constituent atoms are in the gas phase (W00). In this case the distinction between gas and dust can be suspended. As discussed in Section \ref{sec:intro} the dust in protoplanetary disks may 
undergo both settling and growth. To accommodate both these phenomena we introduce 
two quantities, $\epsilon$ and $\fb$, encapsulating dust settling (the physical removal of dust) and 
grain growth (self-blanketing) respectively. These are grain-specific parameters, and as such 
only affect the contribution due to dust. Therefore we write the total cross-section as, 

\begin{equation}\sxtot=\sxgas+\epsilon\fb\sxdust. \label{eq:gas_dust_combine}\end{equation}
This is also true for combining fitting coefficients, 
\begin{equation}
c_{i}^{tot}=c_{i}^{gas}+\epsilon\fb c_{i}^{dust}.\label{eq:coef_combination}\end{equation}

\subsection{Elemental abundances and composition\label{sec:abundances}}

\begin{figure}

\includegraphics[width=0.95\textwidth]{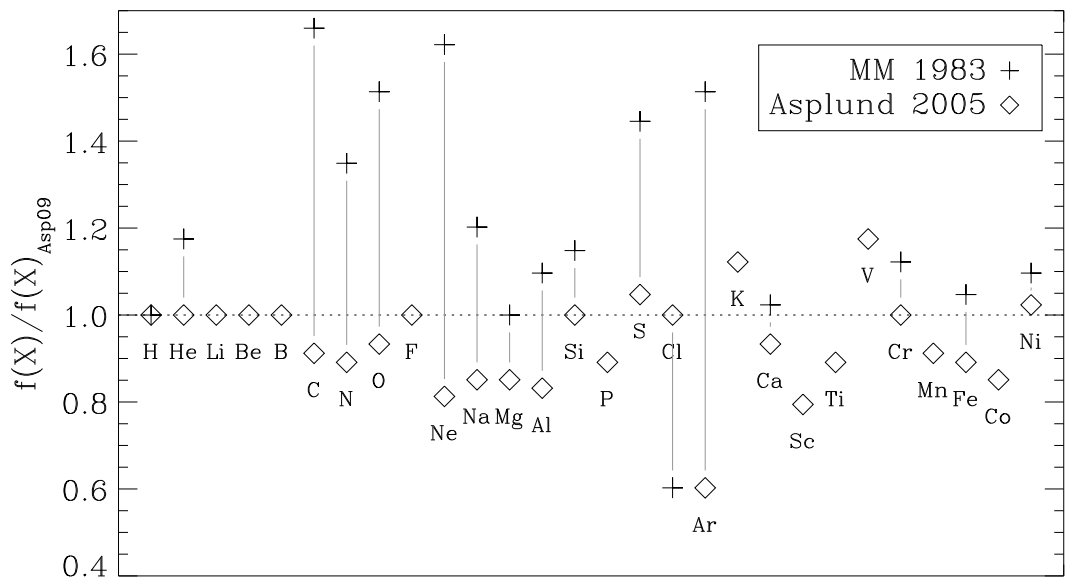}

\caption{Comparison of the solar elemental abundances of \citet{Asplund:2009jl} with those used previously in the literature. The quantity $f(X)$ is the fractional abundance of element X relative to hydrogen nuclei. \textit{Diamonds} - abundances adopted by \citet{Asplund:2005xw}. \textit{Plusses} - abundances adopted by MM83. Of particular note are the variations in the carbon and oxygen abundances, which over time have varied by $>50$\% relative to the \citet{Asplund:2009jl} values.\label{fig:abundances}}
\end{figure}

The elemental abundances we have adopted are given in Table \ref{table_abundances} \citep[based on data from ][]{Asplund:2009jl}. It is from these abundances that we make up our total gas-dust mixture. A comparison with previous estimates of the solar elemental abundances is given in Figure \ref{fig:abundances}. Although all the elements listed appear in the X-ray cross-section to some extent, carbon and oxygen (important contributors at $E\sim1$keV) both have solar abundance measurements that have varied 
considerably over time \citep{Asplund:2005xw}. The distribution of elements between dust 
and gas phases is treated in a largely binary fashion; the metals are entirely incorporated 
into dust $(\beta=1)$, whereas H,He and the noble gases remain entirely in the gas phase 
($\beta=0$). The inclusion of Noble elements in the gas phase is motivated not only by observations and modeling \citep{Pascucci:2007uq,Glassgold:2007fk}, but also by their known low chemical reactivity and low propensity for freezing-out onto grain mantles \citep{Charnley:2001kx}. The exceptions to the simple partitioning of elements are carbon and oxygen, which are divided between the gas and 
solid phases according to \citet{Sofia:2004ud}. Some C and O are necessary in the gas phase to account for observations of circumstellar CO \citep{Calvet:1991vn,Thi:2001zr}.  The partitioning of metals between gas and dust phases is an interesting avenue for further exploration, however we fix it in this paper to limit the number of free parameters \citep[c.f.][]{Balucinska-Church:1992kn}. The total cross-section of a mixture of elements (cm$^2$ per H nucleus) is found by combining the abundances with the atomic X-ray cross-sections $\sigma_Z$ from the online NIST database \footnote{http://www.nist.gov/physlab/data} \citep{Chantler:2000xq},

\begin{equation}\sigma=\Sigma_Z \sigma_Z A_Z\end{equation}

In this paper we only consider neutral species, requiring that the ionization state of the 
gas is low. In order to produce simple fits to the cross-section we neglect solid state features near photoelectric thresholds. The detailed study of X-ray fine structure may provide a useful diagnostic of the composition of solid interstellar material \citep{Lee:2009mw}.  As noted in W00, H$_2$ has a cross-section that is approximately 40\% larger (per H nucleus) than atomic H. We adopt the enhanced H$_2$ cross-section and assume the gas is fully molecular (i.e. $n(\textrm{H}_2 )/n_{\textrm{H}} = 0.5$). Relative to earlier work focusing on the largely atomic ISM, the inclusion of the molecular enhancement increases the cross-section most signiÞcantly at extreme ultraviolet wavelengths where hydrogen opacity dominates.

\section{Results} \label{sec:results}

\begin{figure}

\includegraphics[width=0.95\textwidth]{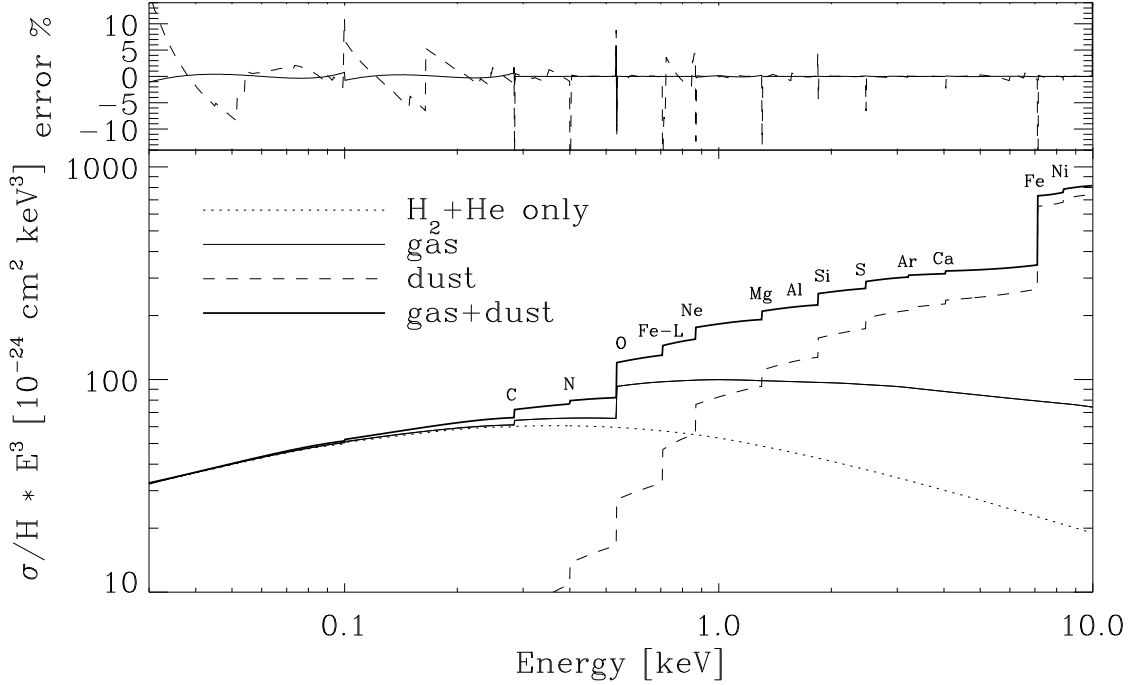}

\caption{\textit{Bottom} - X-ray cross-section of our basic gas-dust mixture using \citet{Asplund:2009jl} 
solar abundances. The parametric fitting coefficients for these curves are given in Table 
2. The Þducial dust contribution is calculated without modiÞcation by grain processes (i.e. 
$\epsilon=\fb=1$). To include the effects of settling and self-blanketing one must combine the gas and dust components following Eqns \ref{eq:gas_dust_combine} and \ref{eq:coef_combination}. The reader should note the $E^3$ scaling - the total cross-section in fact drops with increasing 
energy. \textit{Top} - Residuals of our fitting functions for gas and dust components.\label{fig:cross}}
\end{figure}

The well-tested IDL Þtting routines POLY\_FIT and SVD were used to evaluate and 
test the uniqueness of the Þts. The fitting coefficients are given in Table \ref{table_fits}. The resulting 
opacity curves are shown in Figure \ref{fig:cross}.  By comparison with a gas of pure H and He, Figure \ref{fig:cross} clearly shows that C, O and Noble gases Ne and Ar, contribute signiÞcantly to the gas opacity at energies $E > 0.3$keV.  Away from photoelectric edges the fitting errors are typically at the 1\% level, which is small compared to the compositional uncertainties of gas and dust. Noticeable in the dust component are significant $(\sim10\%)$ error spikes near to metal K-edge thresholds.  These errors (consistent with those in W00) arise from the numerical discretization of the elemental cross-sections, and from the fundamental limitations of a low-order fit. It is important to note that striving for a more accurate polynomial fit at K-shell edges is not entirely meaningful unless we include actual solid-state effects, which in general are too complex for a low order polynomial fit \citep{Draine:2003fk}.

\subsection{Grain growth, $\fb$\label{sec:growth}}

\begin{figure}

\includegraphics[width=0.95\textwidth]{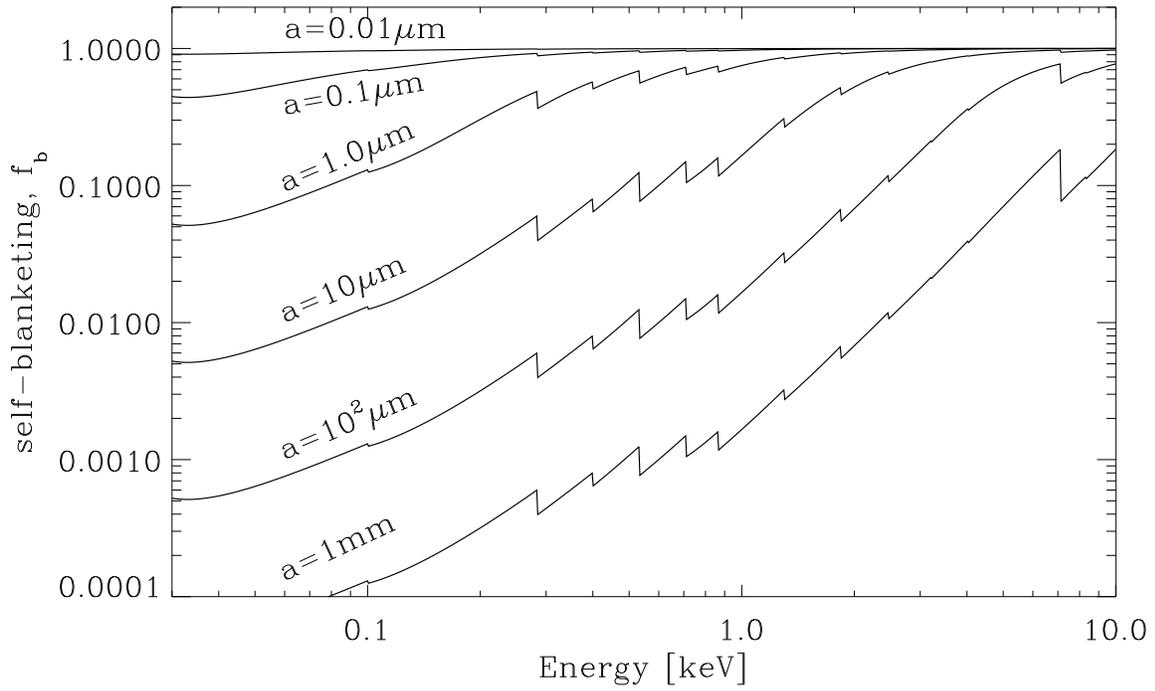}

\caption{Self-blanketing factor for spherical dust grains of radius a. At $E_X=1$keV a dust grain with $a\sim0.5\mu$m has an diametrical optical depth of $\tau\sim1$, marking the onset of appreciable self-blanketing.\label{fig:self_blanket}}
\end{figure}

\begin{figure}

\includegraphics[width=0.95\textwidth]{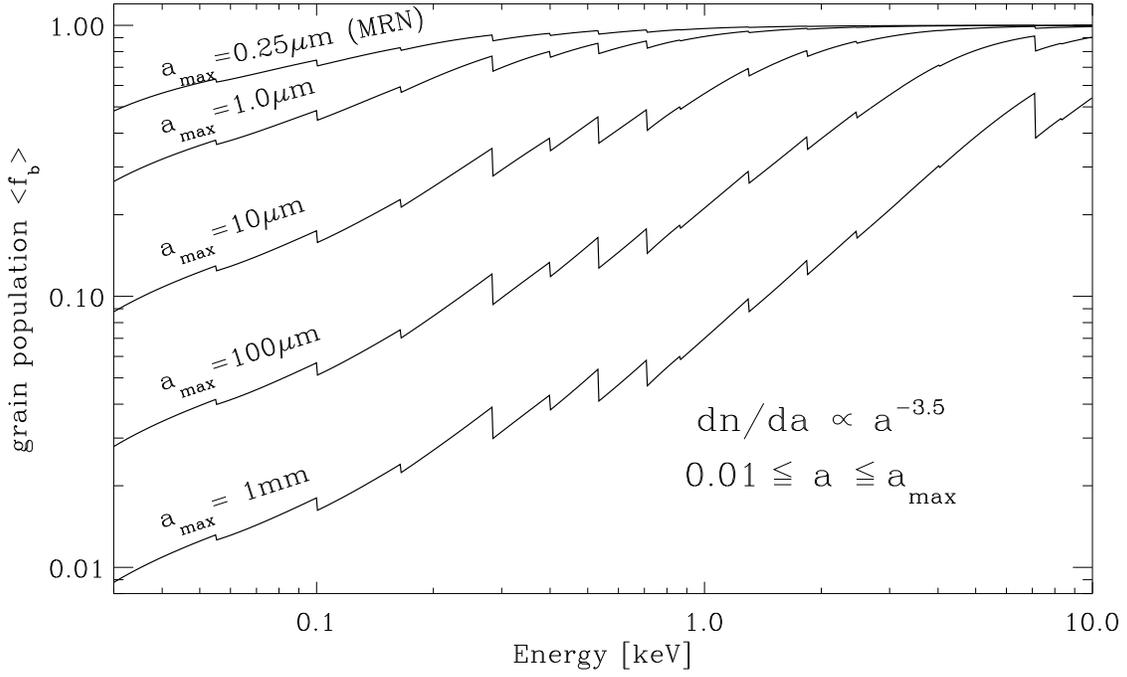}

\caption{Self-blanketing factor averaged over a power-law size distribution of grains. The 
different lines correspond to changing the maximum grain size in the population, $a_{max}$ , 
while keeping the minimum size fixed, $a_{min}=0.01\mu$m. The power-law index of $p=3.5$ and 
$a_{max}=0.25\mu$m is consistent with the classic MRN result for diffuse interstellar dust. \label{fig:effect_of_amax}}
\end{figure}

\begin{figure}

\includegraphics[width=0.95\textwidth]{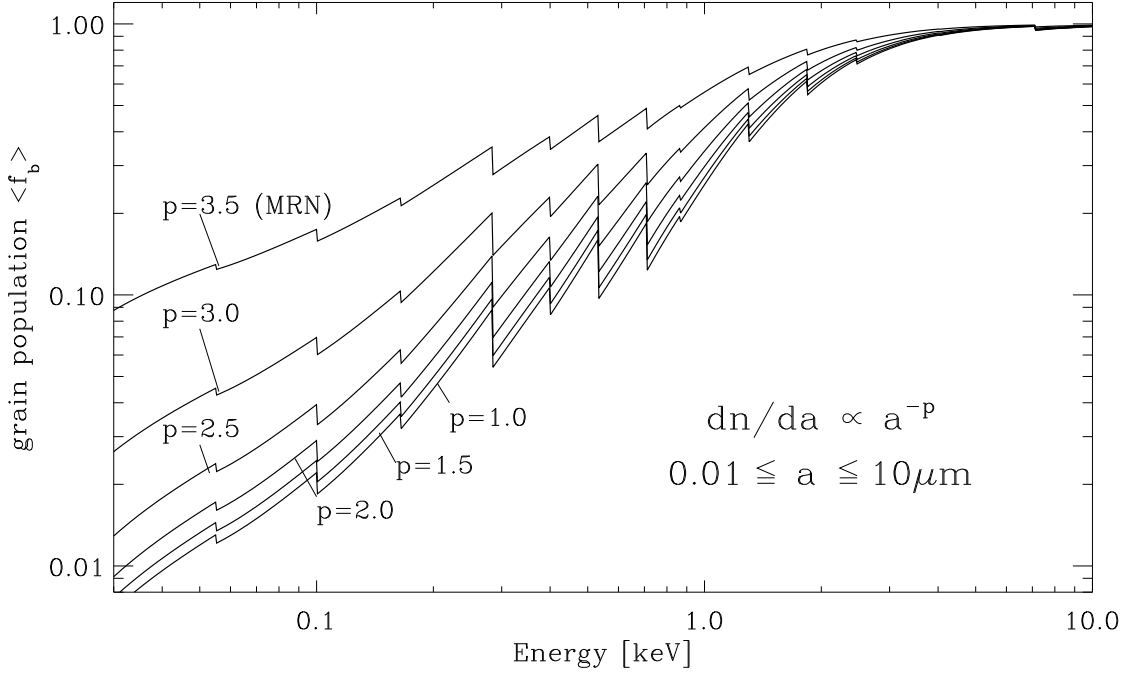}

\caption{Self-blanketing factor averaged over a power-law MRN size distribution of grains. The different lines correspond to changing the power-law index of the population, $p$, while keeping the minimum and maximum grain sizes fixed at $a_{min}=0.01\mu$m and  $a_{max}=10\mu$m respectively. The power-law index of $p=3.5$ is the classic MRN result for diffuse interstellar dust. We have extended $a_{max}$ to a value larger than that speciÞed by MRN, since some grain growth is required before self-blanketing manifests itself. \label{fig:effect_of_p}}
\end{figure}

The derivation of the self-blanketing factor $\fb$ for spherical, homogeneous grains is described in Appendix A.  Plots of $\fb$ versus X-ray energy for a range of grain sizes are shown in Figure \ref{fig:self_blanket}. As a function of energy it bears the imprint of $\sxdust$ since $\fb$ depends upon the optical depth of the grain. At low energies ($E \ll 1$keV) the self-blanketing effect can be very strong (the grains are optically thick, $\fb\ll1$), but it is not seen in the total (gas+dust) opacity because at these energies 
the total opacity is dominated by gas, not dust. At energies $E>1$keV the metals in dust grains 
dominate the total (gas+dust) opacity but tend to be optically thin (for $a<100\mu$m), and 
as a result the self-blanketing effect is weak. At intermediate energies ($E\sim1$keV) the self-blanketing factor departs from unity once grains have grown to $a\sim0.5\mu$m. Grains as 
large as $a\sim1$mm are extremely optically thick, hiding 99.9\% of their constituent atoms 
from the ambient X-rays. 
The grain-size distribution in protoplanetary disks is largely unknown, although the 
process of coagulation will tend to drive mass into larger grains. The 
ensemble averaged self-blanketing factor, $< \fb >$, for an population of grains with a 
power-law size distribution, $dn/da \propto a^{-p}$, is given by the average, 

\begin{equation}
<\fb>=\frac{\int^{a_{max}}_{a_{min}} \fb a^{3-p}\textrm{d}a}{\int^{a_{max}}_{a_{min}} a^{3-p}\textrm{d}a}.
\end{equation} 
where $p = 3.5, a_{min}\sim0.01\mu m$ and $a_{max}\sim0.25\mu$m for the classic MRN distribution. The mass per decade of grain size in the MRN distribution is weakly dominated by the largest grains. This becomes increasingly true if we flatten the distribution through grain coagulation (taking mass from the small grains and placing it in the big grains). It is therefore of interest to briefly study the effects of changing the maximum grain size $a_{max}$ and power-law index $p$. In Figure \ref{fig:effect_of_amax} we preserve the slope of the MRN distribution but allude to grain-growth by gradually increasing $a_{max}$ from the interstellar value $\sim0.25\mu$m to $1$mm. The increase in $a_{max}$ draws an increasing large fraction of the dust mass into the realm of self-blanketing. A similar effect can be achieved by reducing the power-law index $p$, as shown in Figure \ref{fig:effect_of_p}. In this case the evaluation of $<\fb>$ is increasingly dominated by grains with sizes $a\sim a_{max}$ (we hold $a_{max} = 10\mu$m so that the curves for $p<2$ can be compared to the curve for the single grain size population $a = 10\mu$m in Figure \ref{fig:self_blanket}). Fitting coefficients for the dust component computed using MRN distributions ($a_{max}=1,10,100$ and $1000\mu$m) are given in Table \ref{table_fits_blanket}.   

The effect of self-blanketing on the total (gas+dust) photoelectric cross-section is shown in the bottom panel of Figure \ref{fig:results_grain}. The onset of grain-growth initially affects the cross-section near $E_X\sim1$keV - coincident with the peak X-ray emission from T Tauri systems.

\subsection{Dust settling, $\epsilon$\label{sec:settling}}
The effect of dust settling is shown in the upper panel of Figure \ref{fig:results_grain}, and is straightforward to interpret. To isolate this effect we show its affect applied to the standard interstellar MRN grain size distribution, for which self-blanketing is mild due to the preponderance of small grains ($a<0.5\mu$m).  Once $\epsilon$ has dropped to below $0.1$ (consistent with T Tauri systems in Taurus) the contribution made by dust to the total opacity at $E=1$keV is small compared to that of the gas. The gas sets a basal floor to the cross-section. Further reduction of the cross-section is only possible by depleting gas phase metals and noble elements. 

\subsection{X-ray scattering\label{sec:scattering}}

The primary purpose of this paper is to quantify the photoelectric absorption cross-sections that determine how X-ray energy is deposited into gas and dust.  The more general consideration of how X-rays are transported through space requires the inclusion of scattering.  Restricting the discussion to scattering by spheres of radius $a$, the main parameters that determine scattering efficiencies are the size parameter $x\equiv2\pi a/\lambda$ where $\lambda$ is the X-ray wavelength, and the real part of the complex refractive index, $m$.  \citet{van-de-Hulst:1957ta} provides an authoritative discussion of scattering processes throughout this parameter space (see Fig. 20 Ch. 10). While Mie theory provides a rigorous way to compute the extinction of individual particles, a number of approximations are applicable at the limits of this parameter space.  The region of interest is reduced somewhat since $m-1\ll 1$ for most materials (including interstellar dust candidates) at X-ray wavelengths \citep[$E>0.1$keV,][]{Draine:2003fk}.   A refractive index close to unity implies that the phase of an X-ray inside the grain only slowly deviates from that of an exterior X-ray. When the total differential phase change $2x(m-1)$ is smaller than unity (true for most interstellar grains) one may use the Rayleigh-Gans approximation to compute the scattering efficiency \citep{Mathis:1991oq}.  The Rayleigh-Gans approximation asserts that the grain can be viewed as an ensemble of Rayleigh scattering centers \citep{Overbeck:1965kx}.  The sum total of these fields gives a scattered field that is sharply peaked in the forward direction: the property responsible for the narrow halos observed around X-ray point sources \citep{Rolf:1983ys,Predehl:1995vn,Smith:1998fk}.  When computed for the \citet{Weingartner:2001uq} grain population, $90\%$ of 1keV X-rays are scattered by less than one degree away from the forward direction\footnote{In the limit of geometrical optics (large grain sizes, $2x(m-1)\gg1$) the absorption and scattering (diffraction) cross-sections are each equal to the geometrical cross-section $\pi a^2$ (c.f. extinction paradox).}.

In Figure \ref{fig:scattering} we show alongside the photoelectric absorption cross-section the scattering cross-sections of both gas and dust computed by \citet{Draine:2003fk}. When the $E^3$ factor has been taken into account it can be seen that the non-relativistic (Thomson) scattering by \textit{gas} is essentially wavelength independent.  Thomson scattering - the interaction of a photon with a free (or weakly bound) electron - is due almost entirely to the H and He since these elements carry the majority of the electrons \citep{Igea:1999jo}.   It only contributes significantly to the total extinction at $E\ge6$keV.  Even in dust depleted scenarios, scattering by gas is only ever important for the highest energy photons observed in T Tauri spectra.  In contrast, dust scattering contributes significantly at lower energies, $E\sim1$keV.  However, while small-angle scattering by grains is important for interpreting observations of X-ray halos, the very narrow phase function associated with dust scattering suggests that multiple scattering events are required to produce a significant diffuse X-ray field inside a protoplanetary disk. Thus, while the dust scattering cross-section may be large in magnitude it does not redirect radiation effectively.  Since the total photo-electric absorption efficiency at $E\sim1-2$keV is never much smaller than the scattering efficiency, the majority of X-ray photons emitted by T Tauri stars will have been absorbed before a significant diffuse component is generated.

\section{Summary\label{sec:summary}}

We have used the most recent estimates of the solar abundance to construct simple parametric fitting functions for the X-ray photoelectric cross-sections of material in T Tauri disks. By separating the medium into gas and dust components we have shown how the X-ray opacity of T Tauri disk material will change as dust grows in size or settles towards the disk midplane. Here we provide a summary of our results: 

1. Separate polynomial Þts are presented for the photoelectric absorption cross-section 
of gas and dust (Tables \ref{table_fits} and \ref{table_fits_blanket}, Fig. \ref{fig:cross}). These fits are backward-compatible with MM83.

2. Grains as large as $a\sim1\mu$m are required before the effect of self-blanketing is seen. This effect is most readily seen at $E\sim1-2$keV (Fig. \ref{fig:self_blanket}) - coinciding with the peak X-ray emission of T Tauri systems. Due to the large fraction of small grains ($a<<1\mu$m), considerable grain-growth is required ($a_{max}>>1\mu$m) before the MRN grain-size distribution exhibits appreciable self-blanketing at $E\sim1$keV (Fig. \ref{fig:effect_of_amax}). 

3. Provided sufficiently large grains are present, the self-blanketing factor averaged over a 
power-law grain-size distribution is sensitive to the power-law index (Fig. \ref{fig:effect_of_p}). For size 
distributions flatter than MRN (i.e. $p<3.5$) we observe a stronger ensemble-averaged 
self-blanketing effect, eventually asymptoting to that of a single-sized grain population 
with $a\sim a_{max}$. 

4. Realistic structural and compositional inhomogoneities of grains will change the 
self-blanketing results. Non-spherical grains will self-blanket less effectively. The 
layering of grain mantles may either increase or decrease the self-blanketing factor, 
depending upon the composition and order of these layers. 

\begin{figure}

\includegraphics[width=0.95\textwidth]{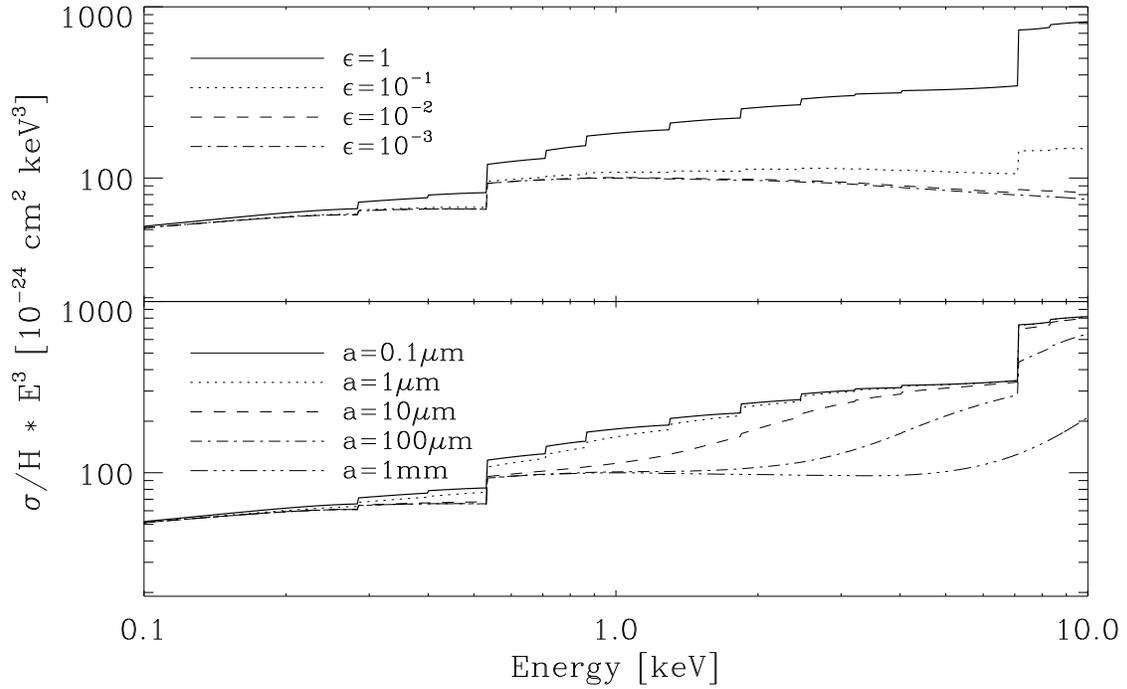}

\caption{Effects of grain processes on the total (gas+dust) X-ray photoelectric cross-section. \textit{Top} - the contribution to the X-ray opacity made by grains varies in proportion to changes in the dust:gas mass ratio, $\epsilon$, occurring as a result of dust settling. \textit{Bottom} - grain growth affecting the dust opacity in a more complicated, energy-dependent manner.  In both plots the gas opacity sets a basal limit.\label{fig:results_grain}}
\end{figure}

\begin{figure}

\includegraphics[width=0.95\textwidth]{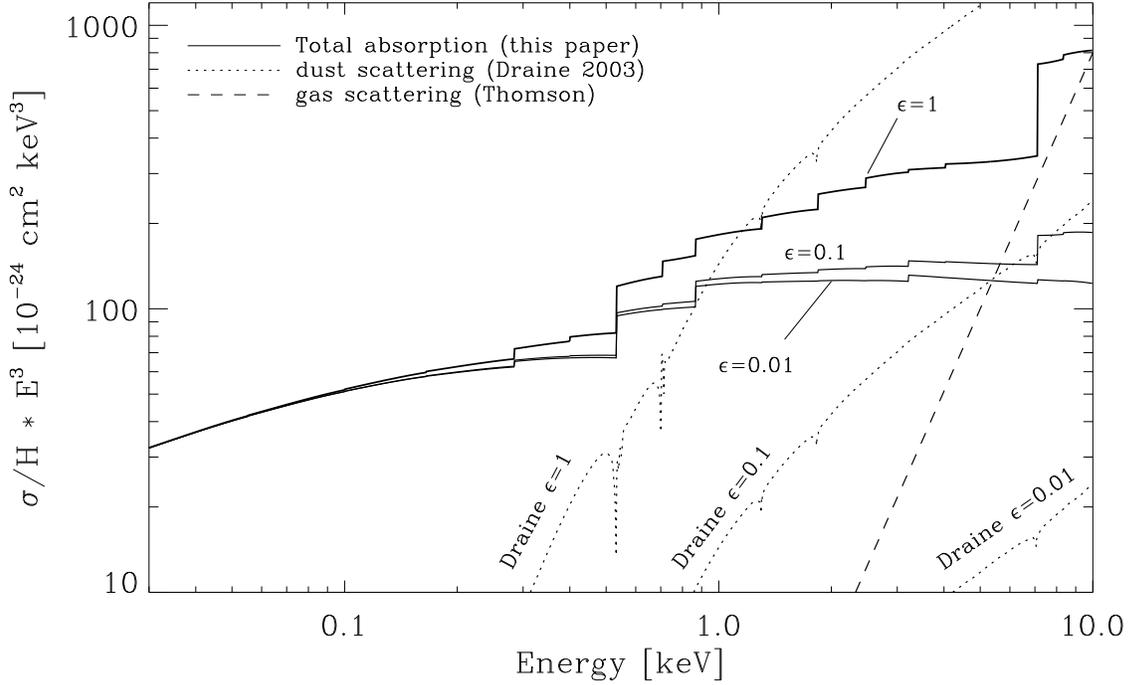}

\caption{Comparison of the (gas+dust) photoelectric absorption (\textit{solid lines}) and scattering 
due to gas (\textit{dashed line}) and dust (\textit{dotted line}), as a function of the settling parameter $\epsilon$. At $E\sim1$keV scattering by gas contributes negligibly to the total cross-section, whereas dust scattering is of the same order of magnitude as dust absorption. At $E_X\sim10$keV the scattering opacity is large with both gas and dust contributing significantly.  Although the gas may play a role in scattering X-rays at these energies, the total photoelectric absorption is dominated by the metals in dust.  Note the $E^3$ scaling - the (Thomson) gas scattering opacity is in fact almost independent of energy\label{fig:scattering}}
\end{figure}

5. The degree of dust settling evident in protoplanetary disks, $\epsilon\sim0.01$, is more than 
sufficient to reduce the X-ray opacity of dust to levels that are small compared to 
those of the gas (c.f. Fig. \ref{fig:results_grain}).  The combined grain-growth and settling observed in protoplanetary disks suggests that X-ray energy in T Tauri systems is preferentially injected into the gas component.  

6. While both grain-growth and settling reduce the X-ray cross-section, the remaining 
gas opacity sets a lower limit that ensures $\sxtot$ ($E\sim1$keV) does not drop below about 50\% of the diffuse ISM value (Fig. \ref{fig:results_grain}).
 
7. Further reduction of the X-ray cross-section in dust-evolved systems is contingent 
upon changes in the metal composition of the gas. The removal of CO and Ne as the 
gas disk dissipates could potentially reduce $\sxtot$ by a further factor of three. 
%8. The albedo, ?, typically increases as the dust evolves towards $\epsilon<0.1$ . At $E_X=1$keV 
%the albedo is always small, ??0.007?0.01, however at $E_X?10$keV it can be very large, ? = 0.5 
%? 0.9 (Fig. X). As such, grain processes will inßuence the character of 
%X-ray transport within disks, especially during the energetic stellar ßaring events that 
%produce 10keV X-rays. 

X-ray counterparts have been identified for many protoplanetary disk systems in Orion, several of which display X-ray spectra consistent with absorption by surrounding material \citep[e.g. COUP][]{Getman:2005ve,Kastner:2005ly}. By comparing the absorption column inferred from X-rays to that determined from visual extinction (by dust), it is in principle possible to place constraints on the composition and dust-to-gas ratio of intervening disk material.  Edge-on sources COUP 419 and COUP 241 are such examples, suggesting that the absorbing material is considerably deficient in dust (an order of magnitude or more relative to to interstellar value).

\citet{Gudel:2008zr} have detected bipolar X-ray jets emanating from the strongly accreting T Tauri system DG Tau.  Comparison of the two jets indicates that one is probably viewed through an excess column of material ($N_H\approx2.7\times 10^{21}$cm$^{-2}$) due to the surrounding disk.   Unfortunately, while this scenario potentially affords a differential analysis of the intervening material, the visual extinction along this line of sight is uncertain, rendering the gas-to-dust ratio determination inconclusive.

Far-ultraviolet radiation (FUV) is another important source of energy in protoplanetary 
disks. The transport of continuum-FUV is largely controlled by dust, and as such is highly 
sensitive to dust evolution. In particular, as dust settles towards the disk midplane the 
upper parts of the disk become increasingly transparent to FUV photons, reducing the FUV opacity to X-rays levels. In contrast, the propagation of X-rays is impeded almost entirely by gas after only a modest degree of dust evolution ($\epsilon\sim0.1$). The deposition of X-rays with energies typical of the steady coronal emission from T Tauri stars ($E\sim1-2$keV) will be deposited in the upper layers of the disk - the 
so-called warm molecular layer. In the models of \citet{Aikawa:2006uq} the warm layer 
of gaseous CO at R=200AU has a typical vertical column density of $N_Z$(CO)$\sim10^{18}$cm$^{-2}$. We expect this layer to be very optically thick, $\tau_{1keV}\sim100$, when viewed at the small 
grazing angles ($\sim5$ degrees) appropriate for impinging stellar X-rays. The impenetrability of this layer 
is mitigated somewhat at higher energies ($E\sim10$keV) where the photoelectric cross-sections becomes
small compared to the scattering cross-section. In this case, Thomson scattering by gas can be an effective means of disk penetration \citep{Igea:1999jo}.

\acknowledgments

TB and EB gratefully acknowledge funding by NASA under grant NN08 
AH23G from the ATFP and SSO programs.


%% in parentheses, after the keyword, but they will not be verified.



\appendix

\section{Derivation of dust self-blanketing factor $\fb$\label{sec:appendix}}

A single interstellar dust grain of sufficient size may become optically thick to X-rays. In such a case the interior atoms see a reduced flux of X-rays relative to the atoms on the surface of the grain. The total absorptive efficacy of the dust grain is therefore less than that of an equal mass of compositionally identical gas. This effect is referred to as self-blanketing, and to calculate its magnitude we must consider the dust grain as a radiative transfer problem by itself (albeit a highly simplified one). In this Appendix we consider the self-blanketing of spherical dust grains - not because the sphere is a physically sensible model for interstellar grains but because the sphere is the most self-blanketing geometry (under isotropic illumination). The self-blanketing effect of real (i.e. non-spherical) grains will be less than that presented here. We adopt the basic notation of W00, although we differ slightly in our approach by explicitly specifying a grain shape (sphere). 
The average cross-section of the constituent atoms in the solid material is given by 
\begin{equation}
\bar{\sigma}(E)=\frac{\Sigma_ZA_Z\beta_Z\sigma_Z(E)}{\Sigma_ZA_Z\beta_Z}=\frac{\sxdust}{\Sigma_ZA_Z\beta_Z}
\label{eq:app_sigma}\end{equation}
The atomic abundances $A_Z$ and depletion factors $\beta_Z$ are provided in Table 1. In effect Eqn 
A1 simply removes the Ôper hydrogenÕ scaling, replacing it with Ôper grain atom.Õ 
The impingent intensity $I_0$ is partially absorbed by the grain, emerging with an 
intensity $I(x)$ depending upon the offset, $x$, between the ray and the grain center. If the 
grain has a number density n then the ray must traverse an optical depth 
\begin{equation}
\tau(x)=2\bar\sigma n \sqrt{a^2-x^2}.
\label{eq:app_tau}\end{equation}

Averaged over the geometric cross-section of the grain the average emergent intensity is
\begin{equation}
\frac{<I>}{I_0}=\frac{2}{a^2}\int_0^a x \exp \left(-2\bar\sigma n \sqrt{a^2-x^2}\right)\textrm{d}x=\frac{2}{\tau_0^2}\left[1-(\tau_0+1)\exp(-\tau_0)\right]
\label{eq:app_I0}\end{equation}
The final solution is expressed in terms of the diametrical optical depth, $\tau_0=2\bar\sigma n a$. Since spheres look the same from all directions Eqn \ref{eq:app_I0} is already angle-averaged. The fraction of 
radiation removed from the beam is then simply $1-<I>/I_0$. If we ignore self-blanketing 
and assume all the constituent atoms in the grain see the same incident flux of photons, the 
fraction of photons removed is $\bar\tau=2/3\tau_0$. Comparing this with the self-blanketed result 
(Eqn. \ref{eq:app_I0}) yields the self-blanketing factor for a spherical grain, 
\begin{equation}
\fb=\frac{3}{2}\frac{1-\frac{2}{\tau_0^2}\left[1-(\tau_0+1)\exp(-\tau_0)\right]}{\tau_0}
\label{eq:app_fb}\end{equation}
Note that a grain with a small self-blanketing effect has a large $\fb$. In W00 the grain shape 
is not explicitly involved in the calculation. Rather, they characterize the grain solely by a notional optical depth, $\bar\tau$ and their self-blanketing factor is given as, 
\begin{equation}
\fb^W=\frac{1-\exp(-\bar\tau)}{\bar\tau}=\frac{}{}\frac{1-\exp(-\frac{2}{3}\tau_0)}{\tau_0}
\label{eq:app_fbw}\end{equation}
In the last step we have expressed their result using the appropriate mean optical depth 
for a sphere, $\bar\tau=2/3\tau_0$. The difference between the W00 expression and our sphere-speciÞc result (Eqn. \ref{eq:app_fb}) is less than 5\% in the range $\tau_0=1-10$. Both functions have the same 
asymptotic behaviour. For most purposes the computationally simpler W00 expression 
given by Eqn. \ref{eq:app_fbw} will suffice. However, for grain-shapes less regular than the sphere the 
self-blanketing factor should be evaluated in a shape-speciÞc manner. Grain structures with 
low volume illing factors will be less self-blanketing.


%\begin{thebibliography}{}
%\end{thebibliography}
%\bibstyle{aps}
%\bibliography{bib_xray_1}
%\bibliographystyle{apj}
%\bibliography{bib_xray_1}		
%\bibliography{bib_xray_2}	

\clearpage

%% the file, uncomment the \plotone, \plottwo, and
%% If you need a layout that cannot be achieved with \plotone or
%% \plottwo, you can invoke the graphicx package directly with the
%% \includegraphics command or use \plotfiddle. For more information,
%% please see the tutorial on "Using Electronic Art with AASTeX" in the
%% documentation section at the AASTeX Web site,
%% http://www.journals.uchicago.edu/AAS/AASTeX.
%%
%% The examples below also include sample markup for submission of
%% supplemental electronic materials. As always, be sure to check
%% the instructions to authors for the journal you are submitting to
%% for specific submissions guidelines as they vary from

%% This example uses \plotone to include an EPS file scaled to
%% 80% of its natural size with \epsscale. Its caption
%% has been written to indicate that additional figure parts will be
%% available in the electronic journal.

%====================================================

\begin{deluxetable}{cccc}
\tablecolumns{4}
%\tablewidth{320pt}
\tablewidth{0pc}
\tablecaption{Elemental abundances and fraction in dust \label{table_abundances}}
\tablenum{1}
\tablehead{\colhead{Element} &  \multicolumn{2}{c}{Abundance} & \colhead{Dust Fraction\tablenotemark{1}} \\ 
\colhead{} & \colhead{$12+ \log A_{Z}$} & \colhead{$A_{Z}=n_{Z}/\nh$} & \colhead{$\beta_{Z}$}  } 

\startdata
H  & $12.0$ & $1$ & $0$\\
%\hline 
He & 10.93 & $8.51\times10^{-2}$ & 0\\
%\hline 
Li & 1.05 & $1.12\times10^{-11}$ & 1\\
%\hline 
Be & 1.38 & $2.40\times10^{-11}$ & 1\\
%\hline 
B & 2.70 & $5.01\times10^{-10}$ & 1\\
%\hline 
C & 8.43 & $2.69\times10^{-4}$ & 0.28\tablenotemark{2}\\
%\hline 
N & 7.83 & $6.76\times10^{-5}$ & 1\\
%\hline 
O & 8.69 & $4.89\times10^{-4}$ & 0.50\tablenotemark{2}\\
%\hline 
F & 4.56 & $3.63\times10^{-8}$ & 1\\
%\hline 
Ne & 7.93 & $8.51\times10^{-5}$ & 0\\
%\hline 
Na & 6.24 & $1.74\times10^{-6}$ & 1\\
%\hline 
Mg & 7.60 & $3.98\times10^{-5}$ & 1\\
%\hline 
Al & 6.45 & $2.81\times10^{-6}$ & 1\\
%\hline 
Si & 7.51 & $3.24\times10^{-5}$ & 1\\
%\hline 
P & 5.41 & $2.57\times10^{-7}$ & 1\\
%\hline 
S & 7.12 & $1.32\times10^{-5}$ & 1\\
%\hline 
Cl & 5.50 & $3.16\times10^{-7}$ & 1\\
%\hline 
Ar & 6.40 & $2.51\times10^{-6}$ & 0\\
%\hline 
K & 5.03 & $1.07\times10^{-7}$ & 1\\
%\hline 
Ca & 6.34 & $2.19\times10^{-6}$ & 1\\
%\hline 
Sc & 3.15 & $1.41\times10^{-9}$ & 1\\
%\hline 
Ti & 4.95 & $8.91\times10^{-8}$ & 1\\
%\hline 
V & 3.93 & $8.51\times10^{-9}$ & 1\\
%\hline 
Cr & 5.64 & $4.37\times10^{-7}$ & 1\\
%\hline 
Mn & 5.43 & $2.69\times10^{-7}$ & 1\\
%\hline 
Fe & 7.50 & $3.16\times10^{-5}$ & 1\\
%\hline 
Co & 4.99 & $9.77\times10^{-8}$ & 1\\
%\hline 
Ni & 6.22 & $1.66\times10^{-6}$ & 1\\
\enddata 
\tablenotetext{1}{The fraction of element $Z$ in grains.}
\tablenotetext{2}{\cite{Sofia:2004ud}}
%\tablecomments{fdfdfdf}

\end{deluxetable}

%==================yer basic gas and dust coefficients =========================

\begin{deluxetable}{ccccccccccccc}
\tablecolumns{12}
%\tablewidth{320pt}
\tablewidth{0pc}
\tablecaption{Fitting coefficients for different components}
%\tablenum{1}
\tablehead{\colhead{Energy Range} &  \multicolumn{3}{c}{H$_{2}$+He} &&  \multicolumn{3}{c}{gas} &&  \multicolumn{3}{c}{dust} \\ \cline{2-4}  \cline{6-8}  \cline{10-12}
\colhead{(keV)} & \colhead{$c_{0}$} & \colhead{$c_{1}$}& \colhead{$c_{2}$}&& \colhead{$c_{0}$}& \colhead{$c_{1}$}& \colhead{$c_{2}$}&& \colhead{$c_{0}$}& \colhead{$c_{1}$}& \colhead{$c_{2}$} } 
\startdata
0.030-0.055 & 14.3 & 722 & -4190 && 14.2 & 727 & -4130 && 0.0344 & -1.62 & 88.2\\
0.055-0.100 & 22.3 & 432 & -1530 && 22 & 445 & -1550 && -0.147 & 4.19 & 48.1\\
0.100-0.165 & 31.5 & 246 & -578 && 31 & 263 & -614 && -0.677 & 14.9 & 9.6\\
0.165-0.284 & 42 & 120 & -198 && 43.7 & 112 & -165 && -1.12 & 23.6 & -16.2\\
0.284-0.400 & 51.5 & 50 & -68.7 && 49 & 86 & -103 && 0.188 & 24.6 & -1.09\\
0.400-0.532 & 58.1 & 16.8 & -26.4 && 58.6 & 36.9 & -39.9 && -3.57 & 55.5 & -37.9\\
0.532-0.708 & 63.2 & -2.37 & -8.44 && 48 & 130 & -82.2 && -8.24 & 89.6 & -48.1\\
0.708-0.867 & 66.8 & -12.5 & -1.18 && 77.4 & 46.3 & -22 && 57.1 & -49.9 & 52.1\\
0.867-1.303 & 69.4 & -17.5 & 1.2 && 80.1 & 69.8 & -28.3 && 9.11 & 72.7 & -20.8\\
1.303-1.840 & 71.6 & -22.3 & 3.56 && 117 & 7.43 & -1.87 && -8.71 & 106 & -25.7\\
1.840-2.471 & 66.6 & -16.8 & 2.06 && 107 & 16 & -3.75 && 34.9 & 72.4 & -11.4\\
2.471-3.210 & 61.4 & -12.6 & 1.21 && 106 & 13.6 & -2.63 && 23.6 & 85.1 & -11.3\\
3.210-4.038 & 56.6 & -9.6 & 0.75 && 138 & -1.99 & -0.179 && 116 & 28.2 & -2.55\\
4.038-7.111 & 48.4 & -5.8 & 0.308 && 142 & -4.7 & 0.239 && 191 & -2.92 & 1.09\\
7.111-8.331 & 40.5 & -3.43 & 0.128 && 138 & -3.36 & 0.133 && 812 & -74.7 & 6.49\\
8.331-10.00 & 37.8 & -2.8 & 0.091 && 88.9 & 8.15 & -0.547 && -33 & 137 & -6.39\\
%0.030-0.100 & 18.1 & 550 & -2320 && 17.9 & 560 & -2320 && -0.0359 & 0.873 & 71.4\\
%0.100-0.284 & 36.8 & 169 & -310 && 37 & 175 & -309 && -1.77 & 27.9 & -22.8\\
%0.284-0.400 & 51.5 & 50.4 & -69.1 && 49.3 & 84.4 & -100 && -0.507 & 29 & -8.1\\
%0.400-0.532 & 57.9 & 17.6 & -27.3 && 58.4 & 38 & -41 && -3.63 & 55.8 & -38.1\\
%0.532-0.708 & 63.1 & -2.17 & -8.59 && 49.2 & 127 & -79.2 && -6.53 & 84.2 & -43.8\\
%0.708-0.867 & 66.8 & -12.6 & -1.13 && 78 & 44.7 & -21 && -73.5 & 261 & -132\\
%0.867-1.303 & 69.2 & -17.3 & 1.1 && 79.8 & 70.1 & -28.3 && 9.67 & 71.8 & -20.5\\
%1.303-1.840 & 71.6 & -22.2 & 3.55 && 117 & 7.52 & -1.89 && -14.2 & 113 & -27.8\\
%1.840-2.471 & 66.8 & -17 & 2.1 && 109 & 14.5 & -3.4 && 30.4 & 76.3 & -12.3\\
%2.471-3.210 & 61.5 & -12.7 & 1.23 && 108 & 12.7 & -2.49 && 28.5 & 82 & -10.8\\
%3.210-4.038 & 56.7 & -9.68 & 0.76 && 138 & -2.16 & -0.153 && 118 & 27 & -2.41\\
%4.038-7.111 & 48.4 & -5.82 & 0.309 && 142 & -4.71 & 0.24 && 186 & -1.06 & 0.918\\
%7.111-8.331 & 41.6 & -3.71 & 0.146 && 139 & -3.59 & 0.147 && 1040 & -133 & 10.2\\
%8.331-10.00 & 37.7 & -2.77 & 0.0894 && 95.9 & 6.61 & -0.461 && 31 & 123 & -5.61\\
\enddata 

\tablecomments{The coefficients belong to the fitting function given by Eqn \ref{eq:fn_fitting} and can be combined according to Eqn \ref{eq:coef_combination}.The gas component consists of H, He, Ar, Ne and a fraction $(1-\beta_Z)$ of O and C (see Table 1).  The dust component consists of the remaining elements.  The H+He case is meant to serve as a basal reference.\label{table_fits}}

\end{deluxetable}

%===================dust coefs for various amax (MRN)=================

\begin{deluxetable}{ccccccccccccccccc}
\tablecolumns{16}
\rotate
\tabletypesize{\scriptsize}
%\tablewidth{320pt}
\tablewidth{0pc}
\tablecaption{Fitting coefficients for dust component subject to grain growth}
%\tablenum{1}
\tablehead{\colhead{Energy Range} &  \multicolumn{3}{c}{$a_{max}=1\mu$m} &&  \multicolumn{3}{c}{$a_{max}=10\mu$m} &&  \multicolumn{3}{c}{$a_{max}=100\mu$m} &&  \multicolumn{3}{c}{$a_{max}=1000\mu$m}\\ \cline{2-4}  \cline{6-8}  \cline{10-12} \cline{14-16}
\colhead{(keV)} & \colhead{$c_{0}$} & \colhead{$c_{1}$}& \colhead{$c_{2}$}&& \colhead{$c_{0}$}& \colhead{$c_{1}$}& \colhead{$c_{2}$}&& \colhead{$c_{0}$}& \colhead{$c_{1}$}& \colhead{$c_{2}$} && \colhead{$c_{0}$}& \colhead{$c_{1}$}& \colhead{$c_{2}$}} 
\startdata
0.030-0.055 & 0.0203 & -1.5 & 47 && 0.00785 & -0.576 & 17 && 0.00260 & -0.19 & 5.52 && 0.000829 & -0.0607 & 1.75\\
0.055-0.100 & 0.0192 & -1.64 & 50.9 && 0.0166 & -0.918 & 20.6 && 0.00616 & -0.325 & 6.9 && 0.00203 & -0.106 & 2.21\\
0.100-0.165 & -0.173 & 1.69 & 40.7 && 0.00492 & -0.794 & 21.7 && 0.00703 & -0.371 & 7.65 && 0.00275 & -0.129 & 2.48\\
0.165-0.284 & -1.36 & 15.8 & 2.1 && -0.236 & 2.05 & 14.4 && -0.0475 & 0.28 & 6.03 && -0.0124 & 0.0528 & 2.04\\
0.284-0.400 & -2.96 & 27.5 & -0.943 && -0.61 & 4.29 & 16.7 && -0.103 & 0.539 & 7.52 && -0.0238 & 0.091 & 2.58\\
0.400-0.532 & -9.08 & 61.6 & -37.5 && -3.24 & 17.7 & 2.89 && -0.755 & 3.89 & 4.26 && -0.212 & 1.06 & 1.66\\
0.532-0.708 & -19.2 & 100 & -48.7 && -7.71 & 31.9 & 0.587 && -1.8 & 7.03 & 4.67 && -0.506 & 1.92 & 1.9\\
0.708-0.867 & 17 & 19.2 & 17.5 && 2.13 & 8.01 & 23.4 && 1.92 & -2.27 & 13 && 0.725 & -1.17 & 4.65\\
0.867-1.303 & -14 & 98.2 & -28.7 && -25.5 & 71.1 & -11.5 && -5.4 & 14.6 & 3.75 && -1.4 & 3.74 & 1.93\\
1.303-1.840 & -27.1 & 120 & -28.6 && -63 & 123 & -24.5 && -17.8 & 31.3 & -0.219 && -4.69 & 8.05 & 1.04\\
1.840-2.471 & 15.8 & 84.1 & -13.4 && -59.6 & 119 & -17.9 && -28 & 40.5 & -0.463 && -6.39 & 9.35 & 1.39\\
2.471-3.210 & 10.1 & 91.3 & -12.1 && -63.7 & 121 & -15.4 && -63.4 & 66.4 & -4.23 && -14.7 & 15.2 & 0.66\\
3.210-4.038 & 106 & 31.9 & -2.95 && 40.9 & 56.7 & -5.52 && -71.8 & 72.3 & -5.27 && -15.4 & 15.7 & 0.556\\
4.038-7.111 & 187 & -1.84 & 1.01 && 156 & 6.49 & 0.416 && -1.51 & 43 & -1.96 && -17 & 16.9 & 0.479\\
7.111-8.331 & 800 & -72.4 & 6.37 && 704 & -53.9 & 5.39 && 183 & 37.1 & 0.832 && 24 & 13.7 & 2.21\\
8.331-10.00 & -39.5 & 138 & -6.42 && -90.7 & 145 & -6.69 && -397 & 181 & -7.76 && -276 & 86 & -2.05\\
\enddata 

\tablecomments{The coefficients apply only to the dust component.  $a_{max}$ refers to the maximum grain-size in a MRN grain-size distribution.   The total dust mass is held constant. \label{table_fits_blanket}}

\end{deluxetable}

\end{document}